\documentclass[%
preprint,
showpacs,
 amsmath,amssymb,
 aps,
]{revtex4-1}

\usepackage[dvipdfmx]{graphicx}
\usepackage{dcolumn}
\usepackage{bm}
\usepackage{color}
\usepackage{ulem}
\newlength{\figwidth}
\newcommand{\smb}{SmB$_6$}
\newcommand{\ybb}{YbB$_{12}$}
\newcommand{\TK}{$T_{\mathrm{K}}$}

\begin{document}


\title{
Topological surface conduction in Kondo insulator YbB$_{12}$
}

\renewcommand{\thefootnote}{\fnsymbol{footnote}}
\setcounter{footnote}{3}
\author{Y. Sato$^{1, }$\footnote{Present address: RIKEN Center for Emergent Matter Science (CEMS), Wako 351-0198, Japan}}
\author{Z. Xiang$^2$}
\author{Y. Kasahara$^1$}
\author{S. Kasahara$^{1, }$\footnote{Present address: Research Institute for Interdisciplinary Science, 
Okayama University, Okayama 700-8530, Japan}}
\author{L. Chen$^2$}
\author{C. Tinsman$^2$}
\author{F. Iga$^3$}
\author{J. Singleton$^4$}
\author{N. L. Nair$^{5, 6}$}
\author{N. Maksimovic$^{5, 6}$}
\author{J. G. Analytis$^{5, 6}$}
\author{Lu Li$^2$}
\author{Y. Matsuda$^1$}

\affiliation{$^1$Department of Physics, Kyoto University, Kyoto 606-8502, Japan}
\affiliation{$^2$Department of Physics, University of Michigan, Ann Arbor, Michigan 48109, USA}
\affiliation{$^3$Institute for Quantum Beam Sciences, Ibaraki University, Mito 310-8512, Japan}
\affiliation{$^4$National High Magnetic Field Laboratory, Los Alamos National Laboratory, MS-E536, Los Alamos, New Mexico 87545, USA}
\affiliation{$^5$Department of Physics, University of California, Berkeley, California 94720, USA}
\affiliation{$^6$Materials Science Division, Lawrence Berkeley National Laboratory, Berkeley, California 94720, USA}

\date{\today}

\begin{abstract}
Kondo insulators have recently aroused great interest because they are promising materials that host a topological insulator state caused by the strong electron interactions.  Moreover, recent observations of the quantum oscillations in the insulating state of Kondo insulators have come as a great surprise.  
Here, to investigate the surface electronic state of  a prototype Kondo insulator \ybb, we measured transport properties of single crystals and microstructures.   In all samples, the temperature dependence of
the electrical resistivity is insulating at high temperatures and the resistivity exhibits a plateau at low temperatures.    The magnitude of the plateau value decreases with reducing sample thickness,  which is quantitatively consistent with  the surface electronic conduction in the bulk insulating \ybb.   Moreover,  the magnetoresistance of the microstructures exhibits a weak-antilocalization effect at low field.  These results are consistent with the presence of  topologically protected surface state, suggesting that YbB$_{12}$ is a candidate material of the topological Kondo insulator.  The high field resistivity measurements up to $\mu_0H=$\,50\,T of the microstructures provide supporting evidence that the quantum oscillations of the resistivity in YbB$_{12}$ occurs in the insulating bulk. 
\end{abstract}


\maketitle

Kondo insulators are a class of strongly correlated electron systems, which  have  a long research history over the past fifty years.  There, hybridization between localized $f$-electron and conduction electron band gives rise to opening of an insulating gap at the Fermi level at low temperatures \cite{Tsunetsugu,Riseborough}.      Clear signatures of a Kondo hybridization gap have been reported in Kondo insulators by various measurements.    Kondo insulators have received renewed interest in recent years \cite{Lu}.   It has been suggested theoretically that some Kondo insulators  are topological insulators which possess topologically protected metallic two-dimensional (2D) surface states.  Until now, topological insulator states have been mostly studied in  non-correlated band-insulators, in which the band inversion is induced by the spin-orbit interactions. 
In Kondo insulators, on the other hand, the band inversion develops through the interplay of strong electron correlations and spin-orbit interactions \cite{Dzero2016}.  Thus  electron correlation effects  are essentially important for the formation of the topologically non-trivial state.   The topological Kondo insulator is a fascinating realization of the 3D topological insulator state in strongly correlated electron systems.   

Among Kondo insulators, SmB$_6$ and YbB$_{12}$ with cubic structures have been most extensively studied.  The electrical resistivity of both compounds increases by several orders of magnitude upon cooling.  However,  at low temperatures,  the resistivity  exhibits a plateau that has no obvious explanation, which  kept puzzling the community.  It has been suggested theoretically  that SmB$_6$ is a strong topological insulator, which exhibits an odd number of surface Dirac modes characterized by a $Z_2$ topological index protected by the time reversal symmetry \cite{Dzero2010}.    Recently the resistivity plateau in SmB$_{6}$  has been attributed to the topologically protected 2D surface state. Indeed, a series of transport measurements in SmB$_6$ provides evidence for the presence of surface states \cite{Kim, Syers, Eo}.   A salient feature of the topological surface metallic state is a helical spin texture due to spin-momentum locking.  In SmB$_6$, such a spin-momentum locking has been directly observed by spin-polarized angle-resolved photoemission spectroscopy (ARPES) \cite{Xu}.   According to the band calculations,  YbB$_{12}$ is a topological crystalline Kondo insulator characterized by a non-zero mirror Chern number protected by the crystalline reflection symmetry \cite{Weng}.  In YbB$_{12}$, however, the nature of the surface metallic state has been little explored, although a surface states has been observed by ARPES \cite{Hagiwara}.






Recently, other salient aspects of SmB$_6$ and YbB$_{12}$ have been reported.  In an external magnetic field, both compounds exhibit quantum oscillations (QOs) despite the opening of significant charge gaps \cite{Li, Tan, Xiang}. As the QOs are a signature of the Landau quantization of the Fermi surface,  the observations have come as a great surprise.  In particular, in YbB$_{12}$, QOs are observed not only in the magnetization (de Haas-van Alphen effect) but also the resistivity (Shubnikov-de Haas,  SdH, effect).    The 3D character of the  SdH signals in YbB$_{12}$ suggests  that the QOs arise from the electrically insulating bulk \cite{Xiang, Xiang2021}.    Moreover, in zero field, sizeable linear temperature-dependent terms in the heat capacity and thermal conductivity are clearly resolved in the zero-temperature limit, indicating the presence of gapless fermionic excitations with an itinerant character \cite{Sato}.  What is remarkable is a spectacular violation of the Wiedemann-Franz law, indicating that YbB$_{12}$ is a charge insulator and a thermal metal.      

To explore the peculiar properties of YbB$_{12}$,  we need more detailed information about the electronic structure, particularly the nature of the surface conduction.    In this paper, we report the results of transport measurements on YbB$_{12}$ single crystals and microstructures with controlled thickness. The magnitude of resistivity plateau at low temperatures decreases linearly with the sample thickness, indicating the presence of a surface conduction. Moreover,  in the micostructued device where the surface conduction is dominant, we observe 
the weak-antilocalization (WAL) effect in small magnetic fields, as reported  in other topological insulators.
These results suggest a signature of spin-momentum locking in the surface state due to topological protection. Our results thus provide a supporting evidence of the topologically protected surface state in 
YbB$_{12}$.

High quality YbB$_{12}$ single crystals were grown by the travelling-solvent floating-zone method \cite{Iga1998}. The crystals were cut from as-grown ingots and polished into a rectangular shape (A1, 2, and 3). The  microstructures (B1, 2, and 3) were fabricated by focused ion beam (FIB) method to control the thickness of the single crystals in micron size. 
For the microstructures, we cut a cuboid lamella from A2 single crystal. The edges of the lamella are oriented along [100] direction and its equivalent crystalline axes. This lamella was then 
transferred to a Si/SiO$_2$ substrate and glued using an epoxy. To make electrical contacts, the lamella was sputter-coated by 100\,nm gold, which was partly removed later by FIB etching from the top surface. The images of the prepared microstructural devices taken by scanning electron microscopy are represented in Fig.\,1(a). For B2 and B3 microstructures, the lamella was further patterned into the thin bar shape by using FIB. 
Dimensions of all the samples used in this study are listed  in Table\,I. The sample thickness $t$ is defined as a short side in the cross section rectangle. The electrical contacts were made using silver paste for the single crystals, and by evaporated gold for the microstructures. The resistivity measurements were performed using a standard four-contact configuration. High-field resistivity measurements were performed using a capacitor-driven 65\,T pulsed magnet at the National High-Magnetic Field Laboratory, Los Alamos, and the data were taken via a high-frequency a.c.\,technique with a specialized digital lock-in program. The driving signal ($f=75$\,kHz) was generated by an a.c.\,voltage source and applied to the sample across a transformer. The current through the sample was monitored using a 1\,k$\Omega$ shunt resistor, and was determined to be 12.4\,$\mu$A. 

\begin{table}[b]
\caption{
Dimensions [length ($\ell$), width ($w$), and thickenss ($t$)] and fitting parameters obtained by using Eq.(1) of the single crystals (A1-3) and microstructures (B1-3). 
}
\label{table}
\centering
\begin{center}
\scalebox{0.9}[0.9]{
\begin{tabular}{ccccccccc}\hline\hline
\multicolumn{1}{c}{\ \ Sample\ \ } 
& \multicolumn{3}{c}{dimension (mm)} & \multicolumn{1}{c}{\ \ \ \ }\\ 
\cline{2-4}
&$\ell$&$w$&$t$&\,$\rho^1_b$ (m$\Omega$cm)\,&\,$\rho_b^2$ (m$\Omega$cm)\,&\,$\Delta_1$ (meV)\,&\,$\Delta_2$(meV)&\,\,$\rho_n$ ($\Omega$cmK)\, \\ \hline
A1 & 
6.84 & 0.52 & 0.51 &0.0740&1.30&11.2&4.23&0.0274\\
A2 & 
1.21 & 0.95 & 0.20 &0.0745&1.25&11.1&4.14&0.0436\\
A3 & 
1.78 & 1.82 & 0.13 &0.0781&1.72&11.0&3.50&0.117\\
B1 & 
\ $37.5\times10^{-3}$\  & \ $19.1\times10^{-3}$\  & \ $8.2\times10^{-3}$\  &0.0910&1.71&10.9&4.15&1.96\\
B2 & 
\ $8.45\times10^{-3}$\  & \ $2.6\times10^{-3}$\  & \ $1.0\times10^{-3}$\  &0.0958&2.90&11.2&3.50&1.96\\
B3 & 
\ $8.45\times10^{-3}$\  & \ $1.0\times10^{-3}$\  & \ $0.6\times10^{-3}$\  &0.0926&3.01&11.1&3.33&2.01\\	
\hline\hline
\end{tabular}
}
\end{center}
\end{table}

\begin{figure}[t]
\begin{center}
\includegraphics[width=0.5\linewidth]{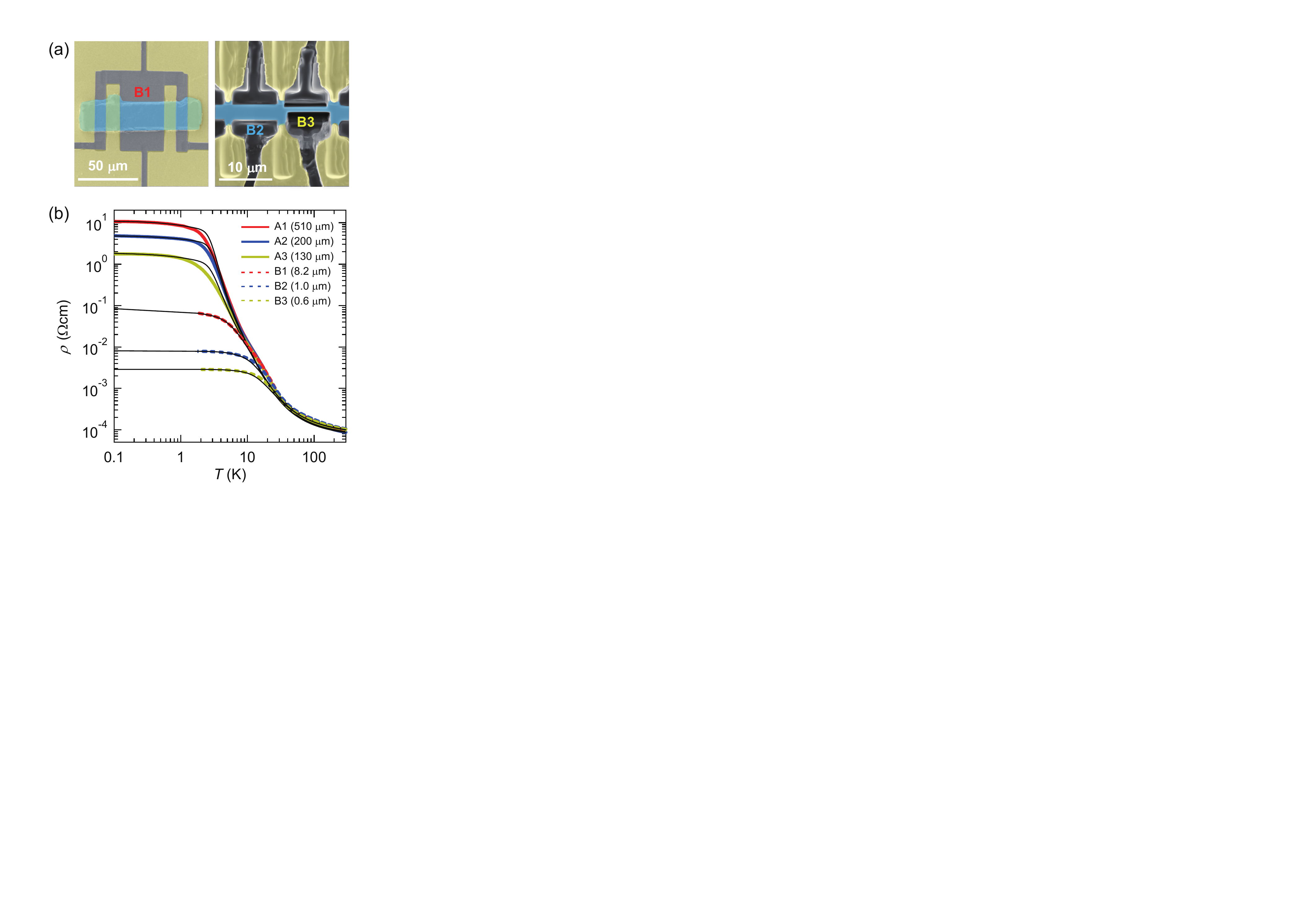}
\end{center}
\caption{
(a) Scanning electron microscope images of microstructures fabricated by FIB. YbB$_{12}$ crystal, gold contacts, and Si/SiO$_2$ substrate are colored in blue, yellow, and grey, respectively. 
(b) Temperature dependence of resistivity $\rho$. The solid and dotted lines are the results of single crystals (A1-3) and microstructures (B1-3), respectively. The black curves are the fits by Eq.\,(1). 
} 
\label{Fig1}
\end{figure}

Figure\,1(b) depicts the temperature ($T$) dependence of the resistivity $\rho$ of the YbB$_{12}$ single crystals and microstructures plotted on a log-log scale. In all crystals and microstructures, $\rho(T)$ increases by several orders of magnitude from room temperature with decreasing $T$, which is attributed to the bulk insulating channel. On further lowering the temperature, $\rho(T)$ becomes weakly $T$-dependent and shows a resistivity plateau, as observed in SmB$_6$ \cite{Kim, Syers, Eo}. Although $\rho(T)$ above 30\,K well coincides for all samples, the magnitude of the resistivity plateau is strongly suppressed with reducing $t$. These results indicate that the resistivity plateau is attributed to the surface metallic states. 
Figure\,2(a) shows the Arrhenius plot of $\rho(T)$ for A1 single crystal. The slope of the plot changes at around $T\approx14$\,K ($1/T\approx0.7\,$K$^{-1}$), showing a two-gap behavior. Fitting with a thermal activation model of resistivity ($\rho(T)\propto\exp(\Delta/2k_\mathrm{B}T)$), we obtain the gap magnitude of $\Delta_1=$ 11.2\,meV and $\Delta_2=$ 4.23\,meV within the temperature ranges $16\,\mathrm{K}<T<50\,\mathrm{K}$ and $4.5\,\mathrm{K}<T<8.5\,\mathrm{K}$, respectively, in agreement with the previously reported values \cite{Iga1998,Xiang}. 

\begin{figure}[t]
\begin{center}
\includegraphics[width=1\linewidth]{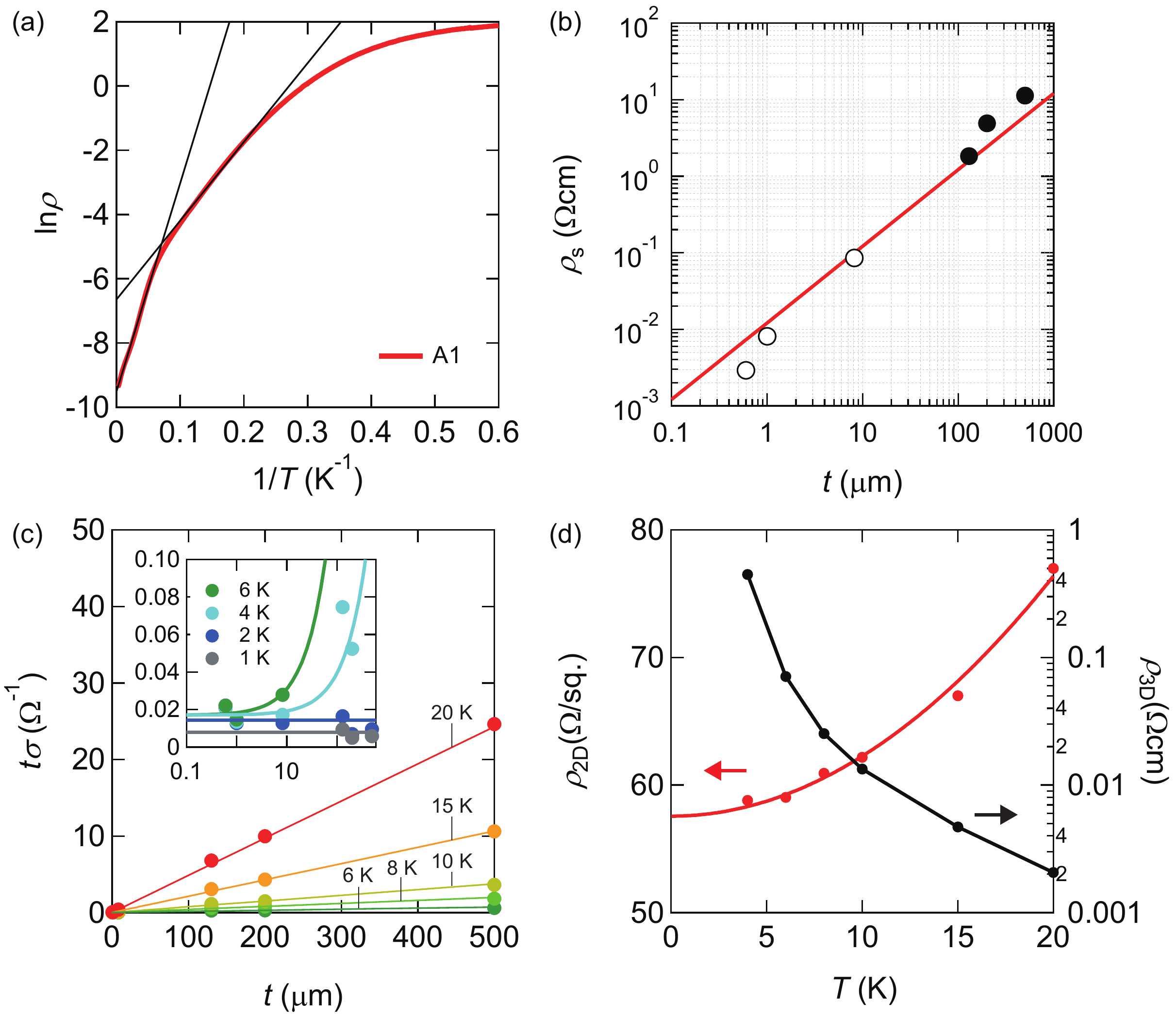}
\end{center}
\caption{
(a) Arrhenius plot of the resistivity $\rho$ for A1 single crystal plotted as a function of $1/T$. The solid black lines are the fits by the thermal activation model. 
(b) Thickness ($t$) dependence of the surface resistivity $\rho_s$ obtained from the fitting of $\rho(T)$ to Eq.\,(1). The filled and open circles represent the results of the single crystals and microstructures, respectively. The red line is a linear ($\rho_{\mathrm{s}} \propto t$) fit.
(c) $t$ dependence of conductance $t\sigma$ with different $T$. The lines are the linear fits. The inset shows the same plot below 6 K in $\log(t)$ scale.
(d) Extracted $T$ dependence of $\rho_{\mathrm{3D}}$ and $\rho_{\mathrm{2D}}$ obtained by the linear fit in (c). The red line shows $T^2$ fit, whereas the black one is a guild to the eyes. 
} 
\label{Fig2}
\end{figure}
\clearpage

A parallel conduction model is used to extract the bulk and surface contributions. The total resistivity is given by 
\begin{equation}
\label{fit}
\frac{1}{\rho} = \sum_{i = 1, 2}\frac{1}{\rho^i_{\mathrm{b}}}\exp{\left( -\frac{\Delta_i}{2k_{\mathrm{B}}T}\right)}  + \frac{1}{\rho_{\mathrm{s}}} + \frac{T}{\rho_{\mathrm{n}}}. 
\end{equation} 
The first term in the right hand side is the contribution arising from the insulating bulk channel with two characteristic gaps $\Delta_i\,(i = 1,2)$ as discussed above. The second term represents  the contribution arising from the surface metallic state. In this model, we first pay attention to its $t$-dependence and we assume that this contribution is temperature independent \cite{Wolgast, Gabani}. The third term is a $T$-linear term, which has been reported in SmB$_6$ and discussed in terms of a possible nodal semimetallic state originating from defects in a crystalline lattice \cite{Harrison,Shen}.  It has been reported that in SmB$_6$, $\rho_{\mathrm{n}}$ is  independent of the sample thickness, indicating the bulk origin \cite{Harrison}.   In YbB$_{12}$, on the other hand,  $\rho_{\mathrm{n}}$ largely depends on the thickness as shown in Table\,I.  Although the origin of $\rho_\mathrm{n}$ is not well understood, $\rho(T)$ for all samples can be well fitted by Eq.\,(1), as shown by the solid black lines in Fig.\,1(b). 
It should be noted that for the bulk channels, the activation gaps $\Delta_i$ as well as the bulk resistivity $\rho_\mathrm{b}^i$ are nearly unchanged by reducing $t$. This demonstrates that the bulk electronic conduction is little damaged during the FIB process, in which a convergent gallium ion beam is irradiated onto a sample and a specific part is made thinner.
In Fig.\,2(b), the magnitude of the surface resistivity $\rho_{\mathrm{s}}$ is plotted as a function of the sample thickness on a log-log scale. As shown by the red line, the surface resistivity increases linearly proportional to $t$.  This indicates that the low-temperature plateau values of the sheet resistance is independent of the sample thickness. 
Thus the present results demonstrate that  
the surface state is an intrinsic property and is robust against the irradiation in YbB$_{12}$. 
We note that in SmB$_6$ ion irradiation does not destroy the surface state but produces a damaged layer that could be poorly conducting \cite{Wakeham}, resulting in a deviation from the linear relationship between $\rho_s$ and $t$. 

The presence of the metallic surface channel is further supported by its extracted $T$-dependence. Sample conductance $t\sigma$ can be expressed by using 2D (3D) resistivity $\rho_{\mathrm{2D}}$ ($\rho_{\mathrm{3D}}$) as, 
\begin{equation}
t\sigma = \frac{1}{\rho_{\mathrm{2D}}} + \frac{t}{\rho_{\mathrm{3D}}}.
\label{separation}
\end{equation}
Figure\,2(c) shows the $t$ dependence of $t\sigma$ with different temperatures obtained from the single crystals (A1-A3) and the microstructures (B1-3). By performing the linear fit, one can separate  $\rho_{\mathrm{3D}}$ and  $\rho_{\mathrm{2D}}$ at a given temperature. We note that we cannot proceed the analysis below $T \sim$ 4 K, because the slope becomes negligibly small with respect to the scattering of the experimental data. Estimation of $\rho_{\mathrm{2D}}$ also becomes difficult when temperature is too high because the change in the second term in Eq.\,\ref{separation} becomes much larger than the first term. Fig.\,2(d) displays the extracted $T$-dependence of $\rho_{\mathrm{3D}}$ and  $\rho_{\mathrm{2D}}$ over the temperature range  4 K $< T <$20 K, which clearly demonstrates the typical transport feature expected in topological insulators: coexistence of 3D insulating bulk and metallic surface. As shown by the red line in Fig.\,2(d), it is remarkable that $\rho_{\mathrm{2D}}$ shows Fermi liquid behavior i.e.,  $\rho_{\mathrm{2D}} \propto T^2$ with a small residual resistivity $\sim$ 58 $\Omega$/sq., which is comparable to that in graphene films grown by chemical vapor deposition \cite{De}.

\begin{figure}[t]
\begin{center}
\includegraphics[width=1\linewidth]{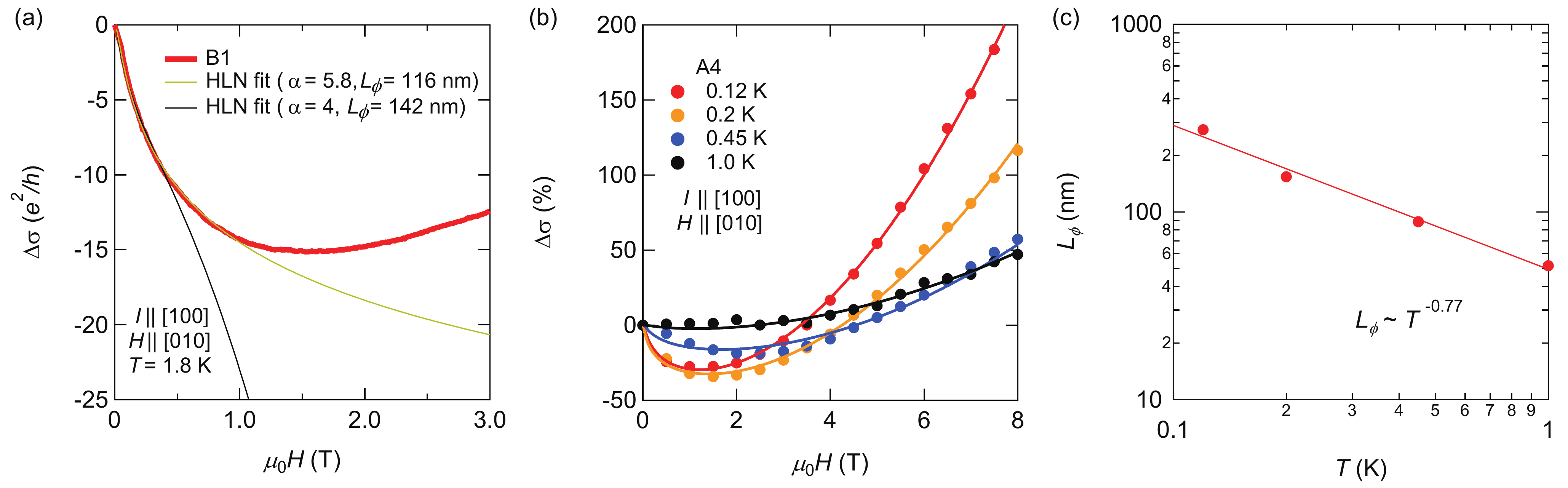}
\end{center}
\caption{
(a) Magnetoconductance $\Delta\sigma = \sigma(H) - \sigma(0)$ at 1.8\,K for the B1 microstructure. The solid thin yellow and black lines show fits to the Hikami-Larkin-Nagaoka (HLN) model (Eq.(3)). In the latter case, $\alpha$ is fixed to 4. (b) Magnetoconductance of the A4 single crystal for different temperatures. The solid lines are the fits to the HLN model. (c) Temperature dependence of the dephasing length $L_{\phi}$ in the A4 single crystal. The line is the fit to $T^{-\gamma}$.
} 
\label{Fig3}
\end{figure}

A signature of the spin-momentum locking provides further support for the topological surface state. 
In the topological surface state,  the spin-momentum locking often gives rise to the WAL effect, which is caused by the  destructive interference between time-reversed electron paths and lowers the sample resistance. This effect is suppressed by applying  magnetic field that breaks time-reversal-symmetry, which causes  a dip in the magnetoresistance (MR) at zero field \cite{Ando,Hasan,Qi}. Such a WAL effect has been observed in SmB$_6$ \cite{Thomas2016} and other topological insulators such as Bi$_2$Te$_2$Se \cite{Assaf}, which is attributed to the topological surface states. 
Figure\,3(a) shows the 2D magnetoconductance $\Delta\sigma = \sigma(H) - \sigma(0)$ in low fields at 1.8\,K for the B1 microstructure in which the surface contribution 
to the electrical transport is larger than the bulk. Magnetic field is applied perpendicular to the major plane of the crystal ($H\,||\,t$ in Fig.\,1(a)) so that the WAL effect  is expected to be most pronounced. With increasing the magnetic field, $\Delta\sigma$ steeply decreases
, which is consistent with the WAL effect. Above 1.5\,T, $\Delta\sigma$ 
turns to increase with $H$.
The positive magnetoconductance at higher fields is a typical feature of Kondo insulators, which is caused by the suppression of the hybridization gap by magnetic field \cite{Xiang, Sugiyama, Shahrokhvand}.

In 2D electronic systems in the limit of long inelastic scattering time, WAL gives a negative quantum correction to the conductivity, which is described by Hikami-Larkin-Nagaoka (HLN) equation \cite{HLN}: 
\begin{equation}
\Delta\sigma = -\frac{\alpha e^2}{\pi h}\left[\psi\left( \frac{B_0}{B} + \frac{1}{2}\right) -\ln\left(\frac{B_0}{B}\right)\right] + \beta B^2,
\label{modified_HLN}
\end{equation}
where $\psi(x)$ is the digamma function, $B = \mu_0H$ is perpendicular magnetic field component to the plane, and $B_0=\hbar/(4eL_\phi^2)$ with the dephasing length $L_\phi$. 
The $B^2$ term includes reduced spin-orbit and elastic scattering corrections as well as the classical cyclotronic contribution \cite{Assaf}. 
Taking into account the contributions from both top and bottom surface side, $\alpha=1$ is expected in a topological insulator with single Dirac band channel. 
 In  the presence of multiple Dirac bands contributing to the surface metallic conduction, the $\alpha$ value is larger than 1.    As four Dirac points appear on the [100] surfaces in YbB$_{12}$ \cite{Weng}, the total number of channels is eight, resulting in $\alpha=4$ if  four Dirac bands individually contribute to the surface conduction with similar $L_\phi$ values.  The black line in Fig.\,3(a) shows the fitting with $\alpha\,=\,4$ and $L_\phi=\,142$\,nm.  The low field data is well reproduced.  The best fit is yielded by using  $L_\phi=116$\,nm and $\alpha=5.8$, as  shown by the yellow line in Fig.\,3(a).  The deviation at high field is due to the negative MR in Kondo insulator.  We note that  in some of 3D topological semimetals and insulators, $\alpha$ obtained from the WAL effect largely deviates from the expected values.  This has been attributed to a contribution from bulk conducting channels and  coherent surface-to-bulk scattering \cite{Pavlosiuk, Sasmal,Thomas2016, Assaf}. 
Thus the present result of $\alpha$ in the YbB$_{12}$ microstructure, which is close to the expected value in the 2D system, indicates that the WAL effect arises from 
the surface state in the presence of spin-momentum locking,
providing supporting evidence of topological surface states in YbB$_{12}$. 

The WAL effect is reproduced even in single crystals as long as they are in the surface conduction regime at low temperatures. Field dependence of magnetoconductance of another single crystal A4 is presented in Fig.\,3(b). The significant drop of the magnetoconductance at low fields is pronounced at low temperatures but it becomes invisible at 1 K. All the data are well fitted by the LHN model up to 8 T as shown by the solid lines in Fig.\,3(b). Temperature dependence of the dephasing length $L_{\phi}$ extracted by the fits is shown in Fig.\,3(c). $L_{\phi}$ decays as $T^{-\gamma}$ with $\gamma$ = 0.77, which is comparable to $\gamma$ reported in SmB$_6$ ($\gamma$ = 0.55) \cite{Thomas2016} and Bi$_2$Te$_2$Se ($\gamma$ = 0.75) \cite{Assaf}. This result suggests that a similar scattering mechanism plays a role in the topologically protected surface states.

\begin{figure}[t]
\begin{center}
\includegraphics[width=1\linewidth]{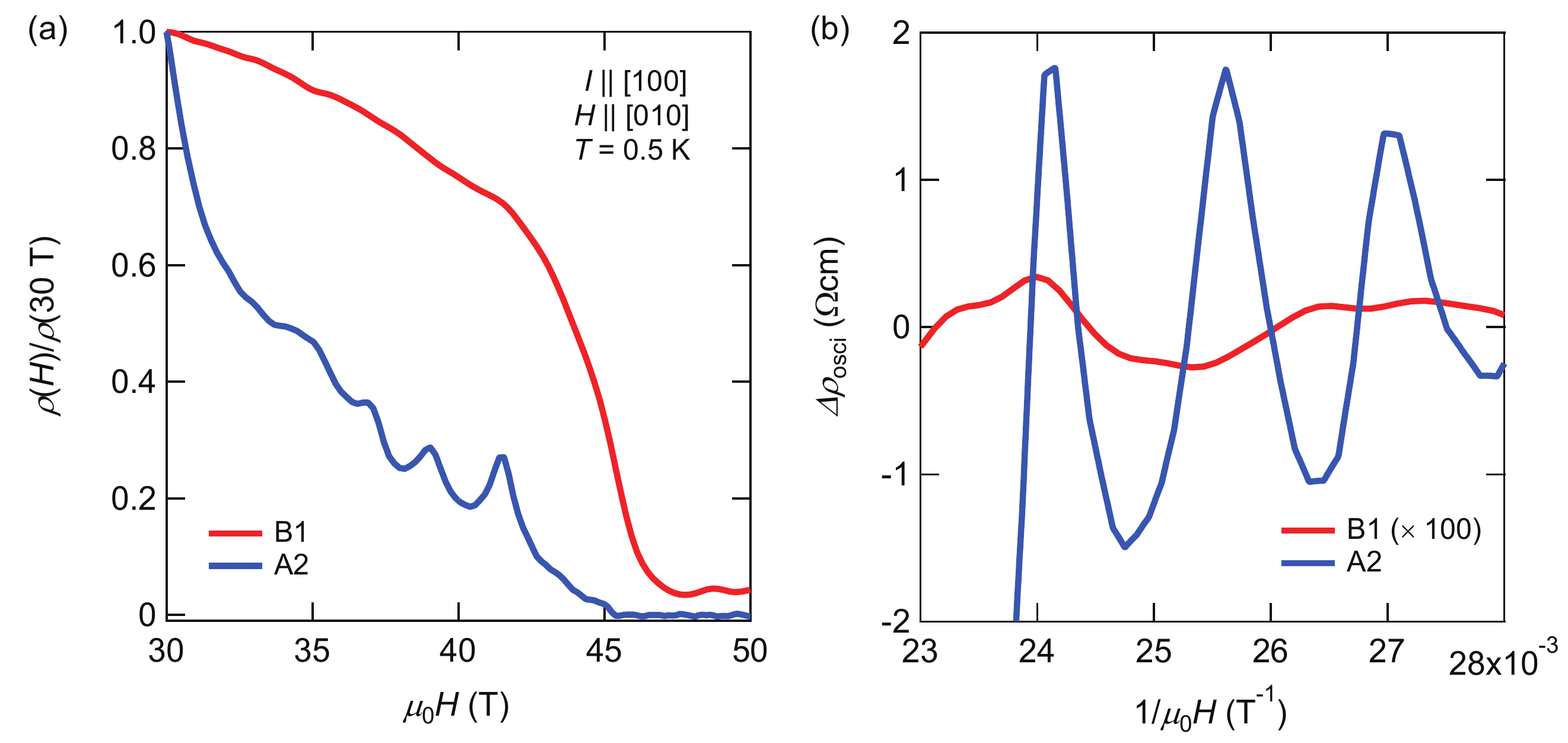}
\end{center}
\caption{
(a) Magnetoresistance normalized by the resistivity at 30\,T, $\rho(H)/\rho(30\,{\rm T})$, measured at 0.5\,K for B1 microstructure (red) and A2 single crystal (blue). (b) The oscillatory part of the magnetoresistance $\Delta\rho_\mathrm{osci}$ obtained by subtracting a polynominal background from $\rho(H)$. 
} 
\label{Fig4}
\end{figure}

To obtain the information on the SdH oscillations reported in bulk YbB$_{12}$ \cite{Xiang}, we measured  the MR of the microstructure at high magnetic fields. 
Figure 4(a) shows the MR of A2 single crystal and B1 microstructure above 30\,T at 0.5\,K. In this field range, the MR of both single crystal and microstructure steeply decreases with increasing $H$ due to the suppression of Kondo gap. The MR displays clear wiggle-like features from about 33\,T to 42\,T in A2 single crystal, while no discernible  wiggle features are observed in B1 microstructure. This can be seen clearly in Fig.\,4(b), which shows oscillatory part of the resistivity $\Delta\rho_\mathrm{osci}$ obtained after subtracting a polynominal background from $\rho(H)$. 
For A2 single crystal, the valleys in $\Delta\rho_\mathrm{osci}$ is approximately uniformly spaced as a function of $1/H$. On the other hand, no periodic oscillation with respect to $1/H$ is observed up to the insulator-metal transition field of $\mu_0H_{\mathrm{IM}}\sim47\,$T in B1 microstructure where the bulk contribution is almost 1/100 times smaller than that in A2. We note that the surface resistance is almost unchanged in single crystals and microstructures, indicating that the scattering rate is comparable in all the samples. Therefore, the SdH oscillations are expected to be observed in B1 microstructure if SdH oscillations arise from the surface conducting channel. 
These results provide supporting evidence that the SdH oscillations in YbB$_{12}$ occurs in the insulating bulk.

In summary, we studied transport properties of Kondo insulator YbB$_{12}$ microstructures fabricated by FIB technique. The low-temperature resistivity plateau decreases with reducing the sample thickness, demonstrating the presence of the surface metallic channels. The MR of microstructures displays a distinct WAL effect,
which is consistent with the spin-momentum locking due to the topological protection. We also present SdH effect is not observed in the microstructures, suggesting 
that unconventional quantum oscillations are inherent properties of the insulating bulk state.

\section*{Acknowledgements}
The work at Kyoto was supported by Grants-in-Aid for Scientific Research (KAKENHI) (Nos. 18H05227, 18H01177, 18H01180 and 18J22138), on Innovative Areas ``Quantum Liquid Crystals'' (No. 19H05824), Overseas Challenge Program for Young Researchers from the Japan Society for the Promotion of Science (JSPS), and JST CREST (JPMJCR19T5).
The work at Michigan was supported by the National Science Foundation under Award No. DMR-1707620 and No. DMR-2004288 (transport measurements), by the Department of Energy under Award No. DE-SC0020184 (magnetization measurements), by the Office of Naval Research through DURIP Award No. N00014-17-1-2357 (instrumentation). A portion of this work was performed at the National High Magnetic Field Laboratory, which is supported by National Science Foundation Cooperative Agreement No. DMR-1644779 and the Department of Energy (DOE).
The work at Berkeley was supported by National Science Foundation under Grant No. 1905397.

\end{document}